\documentclass[]{aa}  
\usepackage{natbib}
\usepackage{graphicx}
\usepackage{txfonts}
\usepackage[colorlinks=true,linkcolor=red,citecolor=blue,urlcolor=blue]{hyperref}
\begin{document}
   \title{Alignment of spiral and elliptical galaxies from Siena Galaxy Atlas with filaments}

   \author{Yuvraj Muralichandran\inst{1,2}, Noam I. Libeskind\inst{2}
          \and
          Elmo Tempel\inst{3,4}
          }

\titlerunning{Alignment of spiral and elliptical galaxies from Siena Galaxy Atlas with filaments}
\authorrunning{Yuvraj M. et al.}

   \institute{Department of Physics and Astronomy, University of Potsdam, 28, Karl-Liebknecht-Straße 24/25, 14476 Potsdam \and
   Leibniz-Institut für Astrophysik Potsdam (AIP), An der Sternwarte 16, 14482 Potsdam, Germany
    \and
    Tartu Observatory, University of Tartu, Observatooriumi 1, 61602 Tõravere, Estonia
    \and
    Estonian Academy of Sciences, Kohtu 6, 10130 Tallinn, Estonia
             }

   \date{Received 12 February 2025; accepted 21 August 2025}
 
  \abstract{The properties of galaxies are known to have been influenced by the large-scale structures that they inhabit. Theory suggests that galaxies acquire angular momentum during the linear stage of structure formation, and hence predict alignments between the spin of halos and the nearby structures of the cosmic web. In this study, we use the largest catalog of galaxies publicly available—the Siena Galaxy Atlas—to study the alignment of the spin normals of elliptical and spiral galaxies with filaments constructed by applying the Bisous process on galaxies ($z \le$ 0.2) from SDSS - DR12. Our sample comprises 32517 disk and 18955 elliptical galaxies that are within 2 Mpc of any filament spine. We find that the spin normals of elliptical galaxies exhibit a strong perpendicular alignment with respect to the orientation of the host filaments, inconsistent with random distributions by up to $\approx$ 13$\sigma$. The spin axis of spiral galaxies shows a much weaker but nonzero alignment signal with their host filaments of $ \approx$ 2.8$\sigma$ when compared with random. These numbers depend on exactly how the significance is measured, as elucidated in the text. Furthermore, the significance of the alignment signal is examined as a function of distance from the filament spine. Spiral galaxies reach a maximum signal between 0.5 and 1 Mpc. elliptical galaxies reach their maximum significance between 0.2 and 0.5 Mpc. We also note that with a tailored selection of galaxies,  as a function of both i) distance from the filaments \& as a ii) function of absolute luminosity, the alignment significance can be maximized. 
  }
\maketitle

\section{Introduction}

    It has been known since at least \citet{Dressler1980} that galaxy properties are related to their environment. A number of studies have shown that a galaxy's environment is correlated not just with its morphology \citep{Kuutma2017,Cooke2023} but also with its star formation \citep{Kauffmann2004,Barsanti2018}, color \citep{Hogg2004,Zehavi2005}, gas content \citep{BarbaraC2013}, metallicity \citep{PengRoberto2013}, and satellite population \citep{2015ApJ...800..112G,2015MNRAS.450.2727T,WangLibeskind2020}, among other characteristics. A number of these correlations are related to the total angular momentum of the stars in a galaxy, namely if it's a rotationally supported disk (possibly with a central bulge component) or if it's a more featureless pressure-supported elliptical galaxy. The connection between environment, angular momentum, and galaxy properties remains an important clue as to how the universe forms galaxies out of collapsing dark matter and gas clouds.

    The correlation between halo shape and the environment (originally dubbed ''cluster alignments``) has been investigated in observations at least since the 1960s \citep{Holmberg1969,Binggeli1982,Cabanela1998}. Numerical simulations followed in a bid to understand the theoretical background as to why massive prolate dark-matter haloes appeared to align themselves with filaments. Theoretical work made a breakthrough in understanding this relationship with the work of \citet{Calvo2007} where a so-called ''spin-transition`` was first noted: low mass haloes tend to have their spins parallel to the filament axes while more massive haloes tend to spin perpendicular to the filament axis. An alignment transition was noted of around halo mass $10^{12} M_{\odot}$. Observational confirmation of such a spin flip has been elusive: Most studies have failed to find it. This is attributed to the fact that the alignment is weak, being only a 10\% effect, owing to projection effects and degeneracies between the inclination angle and the morphology. To date, only a handful of studies have seen a morphological alignment with the large-scale structure. \citet{TempelLibeskind2013} were the first to report that disk galaxies are aligned with their normal axis parallel to the filament spine, while ellipsoidal galaxies are aligned with their long axis parallel to the filament. More recent work by \citet{Codis2012}, \citet{Dubois2014} and \citet{Laigle2014} have confirmed the theoretical prediction for the spin flip. The same trends are confirmed by integral field spectroscopic surveys SAMI \citep{SAMI_BlandHawthorn_2020, Barsanti:2022}, MaNGA \citep{Kraljic:2021}, and CHILES \citep{BlueBird:2020}.

    With the advent of extensive spectroscopic surveys of the local universe, it is possible now to study the correlation between the characteristics of galaxies and their environment in great numbers. Alignment studies are important because they touch on the origin of galactic spin. At least for rotationally supported disks, one would naively expect alignments to be due to their angular momentum acquisition stage.
    
    The most accepted theory to explain how galaxies acquired their angular momentum in their linear regime attributes the protohaloes to acquire their angular momentum to the torque induced by differential tidal forces from their immediate surroundings  \citep{Hoyle1949,Peebles1969,Efstathiou1984,White1984,Porciani2002}. 
    
    Therefore, the orientation of the spin normal of the galaxies is expected to be correlated to the local tidal field, in the absence of any nonlinear evolution \citep{Lee2007}. Since large-scale structures are the representation of complex hierarchical network of matter distribution by tidal shear field \citep{Joeveer1978, Bond1996}, the spin axis of galaxy haloes are expected to show correlation with the large scale structures in their vicinity. Numerous studies have been conducted to visualize this phenomenon, both theoretically and through observations.

    With pure dark matter-only (N-body) simulations, halo spins have consistently shown a mass-dependent 'spin-flip' alignment: low-mass haloes tend to spin parallel to nearby filaments, whereas high-mass haloes rotate perpendicular to the filament direction \citep[e.g.][]{Navarro2004,Calvo2007,Brunino2007,Hahn2007,Codis2012,Libeskind2012,Calvo2013,Codis2018}. This trend was first noted by \citet{Calvo2007} in a N-body simulation run where filaments were identified with a Multiscale Morphology Filter. It was soon corroborated in other N-body studies \citep{Brunino2007,Hahn2007,Codis2012,Zhang2009}, which employed various filament finding techniques but found similar alignment signals. Further analyses with larger simulations and refined cosmic-web classifiers (including tidal-shear eigenvalue methods and the velocity-shear “V-web”) reinforced these results: \citet{Libeskind2012,Libeskind2013a,Libeskind2014,Calvo2013,Trowland2013,Romero2014} all report dark halo spins preferentially parallel to filaments at low masses and perpendicular at high masses. More recent N-body works \citep{WangKang2017,WangKang2018,Veena2018} likewise detect this similar alignment trend, solidifying it as a robust prediction of $\Lambda$CDM gravity-only simulations.

    Hydrodynamical simulations, which follow both dark matter and gas, offer a nuanced view of galaxy spin alignments with filaments, and sometimes disagree. \citet{Navarro2004} ran a high-resolution SPH simulation of a galaxy forming in a super-cluster filament and found that the gas disk can retain its primordial tidal spin even when the dark halo is scrambled. Statistical studies in large volumes, such as Horizon-AGN \citep{Pichon2016,Codis2018} and Illustris-1 \citep{Wang2018}, confirmed a mass-dependent 'spin flip': blue, low-mass galaxies spin parallel to filaments, while massive red ellipticals spin perpendicular. In contrast, SPH-based projects like EAGLE \citep{Veena2019} showed spins preferentially perpendicular at all masses, with no reversal (also seen with MassiveBlack-II \citep{Krowlewski2019}). This diversity underscores how alignment outcomes depend on galaxy mass and the details of hydrodynamics and subgrid physics. An alternative explanation worth mentioning is that filament-finder bias can shift the spin-flip mass: algorithms tuned to thin filaments yield a lower transition mass, so most resolved halos lie above it and show perpendicular spins \citep[see][]{Codis2015}. \citet{Kraljic2020} confirms this by using filament density as a thickness proxy and recovering the expected parallel-to-perpendicular transition from redshift \(z\sim2\) to 0.
    
    Observational studies have also confirmed correlation between the galaxy spin axis with large-scale structure with some key insights \citep[e.g.][]{Kashikawa1992,LeePen2001,Navarro2004, Trujillo2006,Lee2007,Jones2010,TempelLibeskind2013,Tempel2013,Hirv2017,Kraljic:2021,Antipova2025}. Previous observational studies like \citet{TempelLibeskind2013} and \citet{Tempel2013} have verified that this correlation is contingent upon the type of galaxy. Specifically, the spin of spiral galaxies is inclined to align with their nearest filaments, whereas the short axis of elliptical galaxies are perpendicular to the filament. \citet{Zhang2015} reports a mass-dependent correlation in which the spins of spiral galaxies show a faint tendency to be aligned with the intermediate axis of the local tidal tensor. \citet{Pahwa2016}, using the galaxy sample generated from the 2 Micron All-Sky Survey (2MASS) Redshift Survey \citep{Huchra2012}, investigated the alignment between the velocity shear field (V-web, \citet{Hoffman2012}; \citet{Libeskind2013b}) and the galaxy spin. They observed a significant perpendicular signal in elliptical galaxies with respect to the axis of the slowest compression, whereas no such significant signal was detected with spiral galaxies. On the contrary, \citep{Lee2007} report a weak alignment of spiral galaxies with the intermediate axis, mostly attributed to galaxies in dense environments.
    
    Observational results provide us some key acumen about Tidal Torque Theory with vital insights, but often limited by resolution of the galaxies to determine their orientation of the spin normal. With the advent of recent massive spectroscopic surveys, we overcome such limitations. In this study, we present the results of statistical significance of the orientation of spin normals of spiral and elliptical galaxies from Siena Galaxy Atlas with the filaments generated by Bisous process \citep{2016A&C....16...17T}, a marked point process with interactions applied on the galaxy distribution from Sloan Digital Sky Survey. Since filaments are a region where matter and gas falls into, this study would enable us to visualize how galaxy spin is influenced by the local flow of matter/gas and check whether the alignment between the galaxy spin normals and the filaments reflect what is anticipated from the Tidal Torque Theory. The study also provides comparison on how galaxies of different morphology (spiral and elliptical galaxies) are aligned with their host filaments, therefore helps in understanding how large-scale structures influence the angular momentum of morphologically different galaxy systems.   
\section{Data}

 \begin{figure*}[t]
    \centering
    \begin{minipage}{\textwidth}
        \centering
        \includegraphics[width=\textwidth]{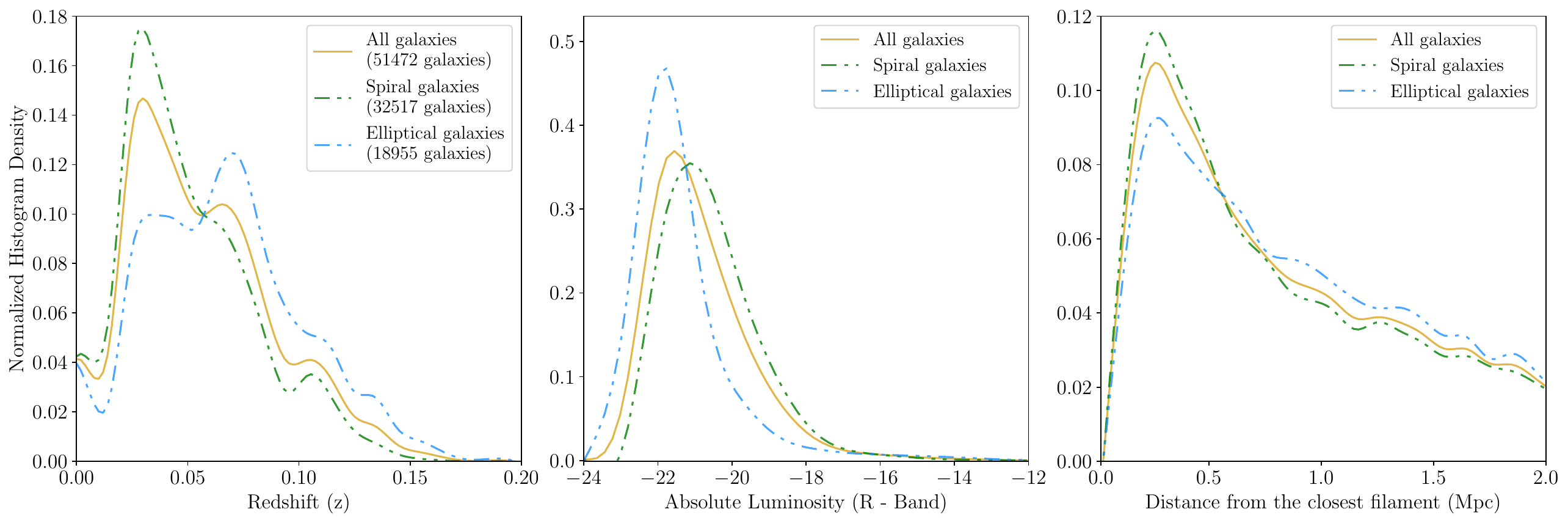}
        
        \caption{Physical properties of the SGA galaxies used in this study. There are a total of 51472 galaxies (32517 spiral galaxies and 18955 elliptical galaxies). The left panel illustrates the (differential) redshift distribution for all galaxies (orange), spiral galaxies (green dotted) and elliptical galaxies (blue dotted). The central panel depicts the absolute magnitude (R-Band) distribution of these galaxies and the right panel illustrates the distribution of galaxy distances from the nearest filament axis (in Mpc).}
        \label{Galaxy Distribution}
    \end{minipage}
\end{figure*}

\subsection{Filaments from Bisous process}
     Although there are many ways of identifying filaments in the distribution of galaxies (e.g. see \citet{Libeskind2018}) we opt to apply an object point process with interactions called the Bisous process \citep{Tempel2014, 2016A&C....16...17T}. The goal of the Bisous filament finder is to model the spines of the filaments in the cosmic web from the distribution of galaxies. The Sloan Digital Sky Survey (SDSS) data release 12 \citep{York2000,Alam2015} is used to construct the filaments. Before filament extraction, we suppress the Fingers-of-God effect to mitigate the redshift space distortions of galaxy groups \citep[see][]{Tempel2016a, Tempel2017}. 
     
     For filament detection in the Bisous model, random segments of thin cylinders are positioned on the distribution of galaxies. Finding a filament is more likely when two such segments are joined and aligned. The morphological and quantitative properties of these intricate geometric objects are determined by following a simple procedure that involves building a model, sampling the probability density that describes the model, and then using statistical inference techniques. To sample the model probabilities, a simulated annealing in conjunction with the Metropolis-Hastings method is employed. The procedure is repeated a numerous times until a network of filaments is formed, each labeled with its coordinates, direction, and statistical significance.
    \citeauthor{Stoica2007} (\citeyear{Stoica2007}, \citeyear{Stoica2010}) provide a thorough explanation of these techniques. 
    
    \citet{Tempel2014} provides a description of the precise implementation of the Bisous method as it is utilized in this investigation. In practice, the approach returns a filament orientation field and a filament detection probability after determining the approximate scale of the filaments. In this model, each discovered structure is by definition a filament, and the relative strength of the structure is described by the detection probability. Filament axes are discovered on the basis of the orientation field and the detection probability field \citep[see][]{2016A&C....16...17T}. Using N-body simulations, it has also been shown that the filaments detected using the Bisous process are well aligned with the underlying velocity field \citep{2014MNRAS.437L..11T}. This emphasizes that the Bisous filaments are well tracing the matter flow in the cosmic web.

\subsection{Siena Galaxy Atlas}
    The sample of galaxies we investigate is taken from the Siena Galaxy Atlas (SGA 2020), which is a comprehensive multi-wavelength optical and infrared imaging atlas of 383,620 nearby galaxies. This atlas is ideally suited to study  the galaxy population as a whole as well as the relationship between and the large-scale structure. Based on the deep, wide-field $grz$ imaging from the DESI Legacy Imaging Surveys DR9 and all-sky infrared imaging from unWISE, the SGA 2020 provides accurate coordinates, multi-wavelength mosaics, azimuthally averaged optical surface brightness, color profiles, integrated and aperture photometry, model images \& photometry and additional metadata for the entire sample \citep{Moustakas2023}. 
    
    Mostly chosen from the HyperLEDA extragalactic collection of known large angular-diameter galaxies, SGA-2020 is the latest edition of the SGA supplemented with additional galaxy catalogs. SGA 2020 is a unique catalog since it includes large angular-size galaxies from HYPERLEDA. It uses multiple sources (DESI Legacy Imaging and WISE) to identify these galaxies, so objects are less likely to be discarded or overlooked due to their size. Therefore, SGA is more complete for large, nearby galaxies, including some large low-surface-brightness galaxies or very nearby giants that could be overlooked in other major catalogs. 
    
    The morphology classification of galaxies are obtained from the HYPERLEDA database (refer: \citet{HYPERLEDA}). The morphological catalog from SGA is not limited only to mere “spiral vs. elliptical” label, but is from a methodical, two‐tiered approach. First, each galaxy inherits a legacy T‐type from the Third Reference catalog of Bright Galaxies \citep[RC3;][]{1991DeVac} as homogenized in HyperLEDA \citep{HyperledaT}. These legacy types are then refined through SGA’s own automated fitting pipeline, which uses deep DESI Legacy grz and unWISE infrared mosaics to perform two‐component bulge and disk decompositions with a flexible Sérsic bulge and an exponential disk. From these fits, SGA derives quantitative structural parameters like bulge‐to‐total light fractions (B/T), Sérsic indices, half‐light radii, and scale lengths for bulge and disk and as well as uniformly measured axis ratios and position angles. 
    
    In addition, SGA computes azimuthally averaged surface brightness and color profiles in concentric elliptical annuli for each band, yielding robust measures of concentration ($R_{90}/R_{50}$), disk scale lengths, and color gradients that can distinguish, for example, a red bulge plus blue disk from a uniformly old stellar population. Because all parameters are extracted by the same code on homogeneously processed mosaics, SGA avoids the systematic biases and cross‐matching complexities that other catalogs like SDSS DR8–based studies would face, such as the reliance on disparate sources such as Galaxy Zoo, NED/RC3 or color/concentration proxies. 
    
    Thus, SGA provides not just a morphological label, but a full suite of well‐calibrated structural diagnostics across tens of thousands of galaxies, enabling the confident selection of pure disks, robust ellipticals, or finely tuned low‐B/T spiral subsamples for environmental and alignment analyses \citep{Moustakas2023}. It is also important to note that this methodology distinguishes late-type spirals from other galaxy types, for example, lenticulars (S0) or early types, and enables us to select galaxies precisely according to their morphology without any contamination.
    
    Together, these features make SGA not just a larger galaxy catalog, but intrinsically more reliable for studying how galaxy spins correlate with the cosmic web. Such an expansive catalog is necessary, especially since in this analysis we have limited our study only to spirals (32,517 galaxies) and ellipticals (18,955 galaxies) that are in the vicinity (within 2 Mpc) of filaments.

    Fig. \ref{Galaxy Distribution} illustrates the distribution of spiral and elliptical galaxies as a function of redshift (left plot), magnitude (middle plot), and each galaxy's distance to its closest filament (right plot).  The distribution of redshift from the left panel of Fig. \ref{Galaxy Distribution}, shows a slight tendency for spiral galaxies to be at lower redshifts compared with the elliptical sample. The middle panel of Fig. \ref{Galaxy Distribution}, shows the absolute magnitude (in R-Band) distribution for both types of galaxies. Although both spirals and ellipticals share similar distributions, the magnitude of elliptical galaxies shows a slightly narrower peak than spiral galaxies. The right panel of Fig. \ref{Galaxy Distribution} indicates that elliptical galaxies are found to be a bit farther away than spiral galaxies from the filament spines; however, the distribution of both galaxy types reaches a maximum around the 0.2 - 0.5 Mpc region from the filaments.

\section{Methodology}
\subsection{Obtaining spin normal for galaxies} \label{sect:spin_normal}
    The SGA provides right ascension (RA), equatorial declination (Dec), magnitude, redshift, axial ratio (b/a), and position angle (PA) of each galaxy either measured based on second moments of the galaxy's light distribution or by the TRACTOR model \citep{Tractor2016}.
    
    Since we are interested in the alignment of a galaxy's spin axis with filaments, we are faced with three complications. Firstly, galaxy catalogs like the SGA provide shapes, not spins. Thus, we must use the short axis of a galaxy's shape as a proxy for its spin. In general, this is a legitimate assumption (and standard practice in the field i.e. \citet{Pahwa2016}) since a number of studies have shown that spin vectors align with the short axes of the stellar distribution of rotationally supported disc galaxies as well as for elliptical galaxies \citep{Cappellari2011,Franx1991,Krajnovic2011}. 
    
    The second issue is that we are only able to measure the direction of projected (2D) axes and not the full 3D axis. In order to estimate the inclination angle of spiral galaxies, and thus the 3D short axis direction, we rely on modeling each galaxy's intrinsic flattening (e.g. see \citet{HaynesGiovanelli1984}; \citet{Lee2007}). Accordingly, the model essentially returns the inclination angle, given the projected short-to-long axis ratio and intrinsic morphology-dependent flattening. \citet{HaynesGiovanelli1984} provides a catalog of intrinsic flattening parameters based on their morphology for all subcategories of spiral galaxies using which the inclination angles are calculated. This inclination is used to de-project the galaxies' normals and results in a 3D vector that can be used to compute an angle with the 3D filament spine direction.
    
    Since we are looking at 3D angles, these will be shown in terms of cosine. It is important to mention that when confronted with an elliptical isophote representing an inclined spiral galaxy, there are two possible inclination angles for a specific spiral galaxy (corresponding to which side of the isophote is closer to the observer). Since there is no (practical) way to break this degeneracy, in this analysis we simply (artificially) double the sample size by including both inclination angles for each spiral. This doubling of the sample size both dilutes the signal and boosts the significance of any signal (by increasing the sample size). These two effects effectively cancel each other out and are common practice \citep[see][]{Pahwa2016,Varela2012,Kashikawa1992}.

    In the case of elliptical galaxies, the inclination angle is assumed to be $90^{\circ}$ (i.e, projected short axis and the spin axis are parallel) as it has been used in previous such studies \citep{Pahwa2016}. Lastly, we note that the ``handedness'' of a galaxy's spin is unobtainable, since all that is able to be measured is the axis of the galaxy's spin axis.

    \begin{figure}[ht]
    \begin{center}
    \includegraphics[scale=0.6]{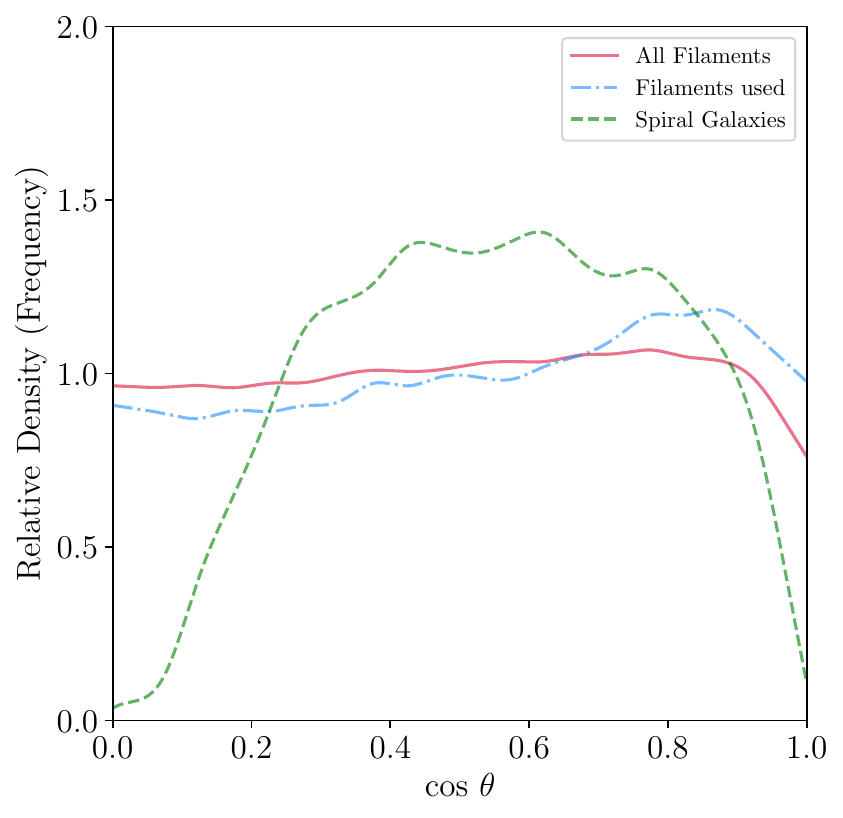}
    \caption{Distribution of the absolute magnitude of the cosine of the angle between the line of sight (l.o.s.) and various objects. In dashed green we show the angle between the l.o.s. and spiral galaxy spin normal. In dot dashed blue with show the angle between the l.o.s and the filaments spines used (namely those filaments with galaxies in the SGA). In solid red we show the angle between the l.o.s. and all the filament points from the Bisous process \citep[see][]{2016A&C....16...17T}.}
    \label{LOS_vs_GalaxiesFilaments}
    \end{center}
\end{figure}
    
    When measuring the alignment between galaxy spin and filament in observational data, we have to be aware of potential biases in the observed data. Both the estimated galaxy spin vector and filament orientation are not uniform with respect to the line of sight. The measured galaxy spin vector is not uniform due to the difficulties of estimating the inclination angle of galaxies. Galaxy filaments are not uniformly distributed due to the redshift space distortions. In Fig. \ref{LOS_vs_GalaxiesFilaments} we show the distribution of measured spin axes and galaxy filaments with respect to the line of sight. Due to the combination of these biases, there will be a measured alignment signal between filaments and galaxy spins even if there is a lack of intrinsic alignment between them. In our statistical analysis (see Sect.~\ref{sec:alignment_significance}) we will take this observational bias into account.

\subsection{Estimation of the significance of the alignment signal}\label{sec:alignment_significance}
   The alignment signal is measured by taking the dot product between the galaxy's implied spin axis and the filament spine (cosine of the angle between the galaxy's spin vectors and the local filament orientation) and constructing a probability distribution function (PDF).
    The resultant distribution is then expressed as a Kernel Density Estimation (KDE) and is tested against the null hypothesis of no (random) alignment. Considering the ambiguity associated with the direction of the spin, i.e, the handedness of the galaxy spin, we use the absolute value of the dot product in constructing the PDF. This has a tendency to weaken any intrinsic signal we find, but the alignment signal that has been obtained here can thus be interpreted as a lower limit, where the actual signal could be stronger than this. All measured signals must be compared to the null hypothesis of a random distribution of angles. Due to implicit biases in observational samples, a random distribution does not necessarily imply a uniform distribution centered around unity \citep{Tempel2013}. As such, the null hypothesis is estimated in the following manner:

    \begin{figure}
    \begin{center}
    \includegraphics[width=\textwidth/2]{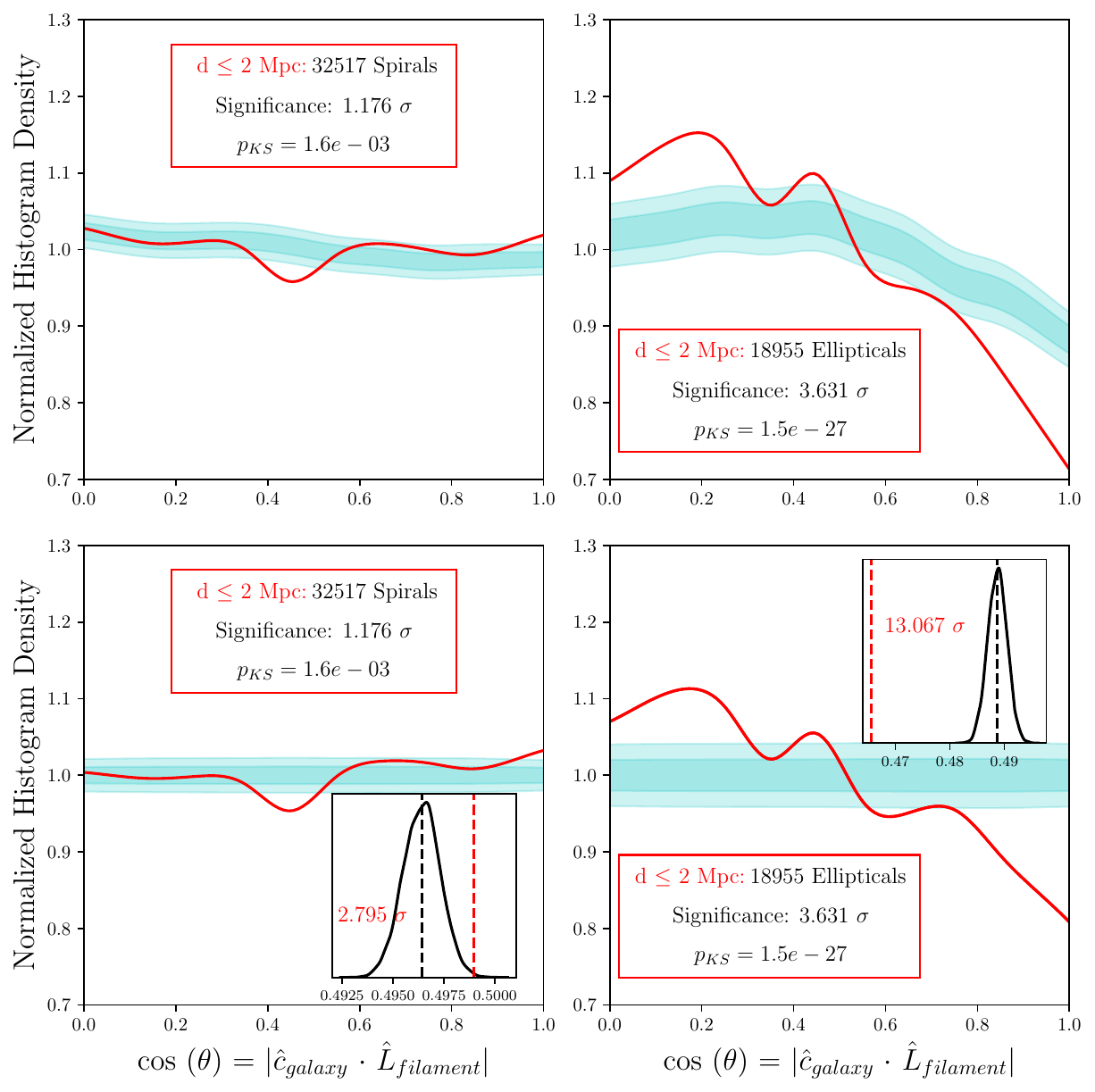}
    \caption{Normalized histogram of cosines of the angles between spines of the filaments and the spin axes of spiral and elliptical galaxies within 2 Mpc (\( \cos \theta = |\hat{c}_{\text{galaxy}} \cdot \hat{L}_{\text{filament}}| \)). The upper panels show the raw alignment signal. The dark cyan band represents the 1 $\sigma$ null hypothesis corridor, beyond which the probability density function is deemed significant, while the light cyan band provides a 2$\sigma$ corridor as a visual comparison. The significance of the alignment signal is expressed in terms of the standard deviations (1$\sigma$) from the random distribution of the null hypothesis. The p-value from the KS test is included to quantify the probability of observing the alignment signal from the given null hypothesis. The lower panels present the alignment signal normalized by the mean random signal such that a random distribution would appear uniform in cos $\theta$. An inset plot depicting the Mean of alignment signal (red line) along with the distribution of the mean from the null hypothesis cases is inserted to visualize how much is the mean of the alignment signal is further from the median of the distribution of the mean from the null hypothesis cases (black line), in terms of standard deviation of the distribution ($\sigma_{\langle \cos \theta \rangle}$). The inset plots also includes the $\sigma_{\langle \cos \theta \rangle}$ values for both distribution, denoted in red text}. 
    \label{Alignment signal - Ellipticals and Spirals}
    \end{center}
\end{figure}

    A randomized control sample is produced by randomizing each galaxy's position angle in the sky, keeping all other properties unchanged. For each randomized sample, a probability distribution of the cosine of the angle between the galaxy's (new, randomized) spin axis and the filament spine to which it is assigned is obtained. This process is repeated 10,000 times. The measured alignment signal can then be compared with the randoms in two ways.
    
    1) The full distribution can be compared. Here we essentially take the mean difference between the measured signal and the median of the 10~000 randoms and express this in terms of the standard deviation of the random sample.
    
2) The mean of the measured signal can be compared with the ``median of the medians'' of the random sample. In other words, each of the 10~000 random distributions has a median angle. This set of random medians itself has a mean value and a standard deviation. The median of the measured signal can then be compared to the mean of the medians of the random values and again expressed in terms of the standard deviation of this distribution.

3) Lastly, a Kolmogorov-Smirnov (KS) test can be applied to ascertain the KS probability that two distributions are drawn from the same parent distribution. Here we apply the KS test to the measured signal and the randoms to see if their cumulative distribution functions are consistent with each other and, if so, at what level of significance. The key advantage of the KS test is its sensitivity to
differences in the entire shape of distributions, making it effective at detecting subtle yet systematic deviations.

We note that these three methods have been utilized in numerous studies in this subfield \citep[e.g.][]{TempelLibeskind2013,Pahwa2016}  
    
\begin{figure*}[t]
    \centering
    \begin{minipage}{\textwidth}
        \centering
        \includegraphics[width=\textwidth]{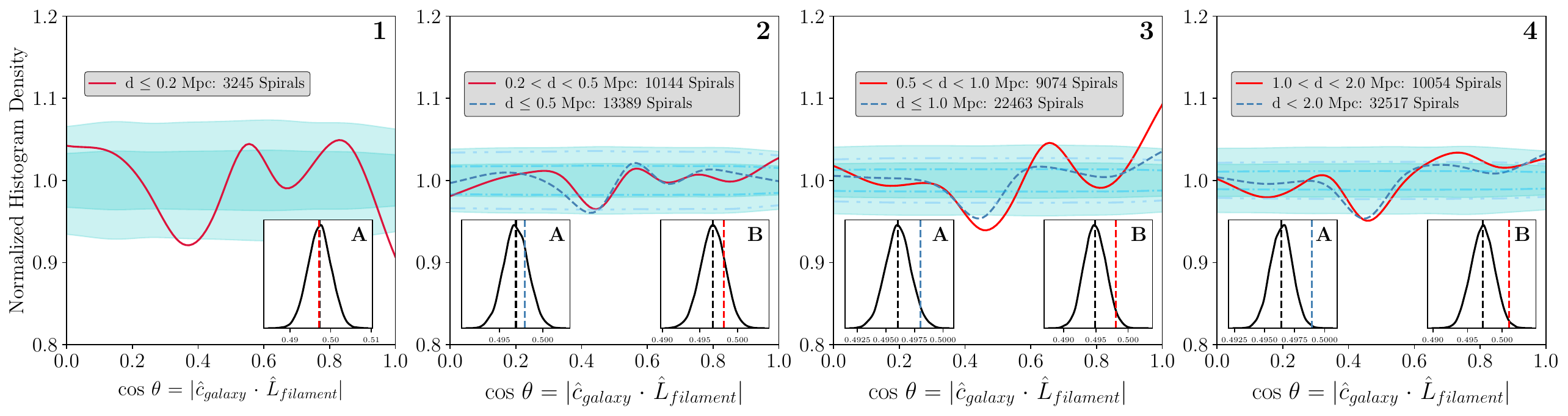}
        
        \caption{Alignment signal - Probability density distribution (PDF) of the cosines of the angle between spines of the filaments and the spin axes of spiral galaxies at different intervals of (< 0.2 Mpc), (0.2 - 0.5 Mpc), (0.5 - 1 Mpc) \& (1 - 2 Mpc, in red). Dark cyan band shows 1 $\sigma$ null hypothesis corridor beyond which the PDF is considered significant. 2 $\sigma$ band (light cyan) is also provided as a visual comparison for the significance of the PDF. For comparison, the cumulative version for the distance bins (i.e, PDF for all the spiral galaxies that are within the distance range of the upper bound of the distance bin) is superimposed (in blue dashed lines) over the PDF for a given distance interval bin along with their error corridors (in blue dash-dotted lines). Inset plots depicting the Mean of alignment signal (red line: for differential subsets and blue line: for cumulative subsets) along with the distribution of the mean from the null hypothesis cases is inserted to visualize how much is the mean of the alignment signal is further from the median of the distribution of the mean from the null hypothesis cases (black line), in terms of standard deviation of the distribution. The statistical significance of alignment in each subset is quantified and presented in Table \ref{TABLE1} (Note: The alignment signal are normalized with the mean of the random signal that the random is based around 1.0 for better inference of the alignment signal, and not under the assumption of null hypothesis to be an uniform distribution (refer Fig. \ref{Alignment signal - Ellipticals and Spirals})) }
        \label{Spiral - Function of distance}
    \end{minipage}
\end{figure*}
        
\begin{figure*}[t]
    \centering
    \begin{minipage}{\textwidth}
        \centering
        \includegraphics[width=\textwidth]{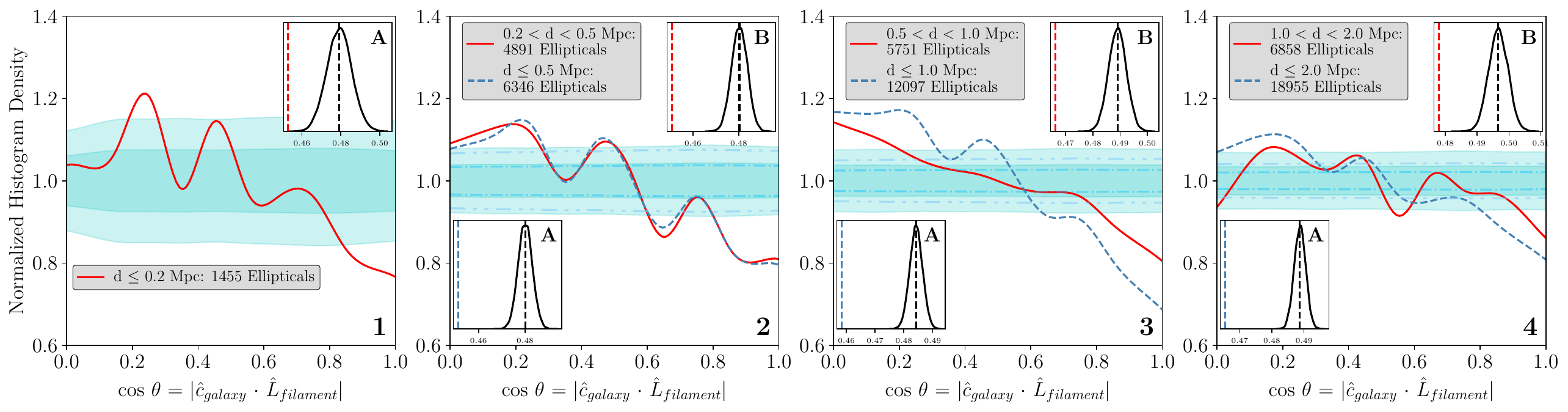}
        \caption{Alignment signal - Probability density distribution (PDF) of the cosines of the angle between spines of the filaments and the spin axes of elliptical galaxies at different intervals of (< 0.2 Mpc), (0.2 - 0.5 Mpc), (0.5 - 1 Mpc) \& (1 - 2 Mpc). Dark cyan band shows 1 $\sigma$ null hypothesis corridor beyond which the PDF is considered significant. 2 $\sigma$ band (light cyan) is also provided as a visual comparison for the significance of the PDF. For comparison, the cumulative version for the distance bins (i.e, PDF for all the elliptical galaxies that are within the distance range of the upper bound of the distance bin) is superimposed (in blue dashed lines) over the PDF for a given distance interval bin along with their error corridors (in blue dash-dotted lines). Inset plots depicting the Mean of alignment signal (red line: for differential subsets and blue line: for cumulative subsets) along with the distribution of the mean from the null hypothesis cases is inserted to visualize how much is the mean of the alignment signal is further from the median of the distribution of the mean from the null hypothesis cases (black line), in terms of standard deviation of the distribution. The statistical significance of alignment in each subset is quantified and presented in Table \ref{TABLE1} (Note: The alignment signal are normalized with the mean of the random signal that the random is based around 1.0 for better inference of the alignment signal, and not under the assumption of null hypothesis to be an uniform distribution (refer Fig. \ref{Alignment signal - Ellipticals and Spirals}))}
        \label{Elliptical - Function of distance}
    \end{minipage}
\end{figure*}

\begin{table*}[t]
    \caption{Statistical analysis of galaxy-filament alignment signals for spirals \& ellipticals within various proximity selections from filament spine}
    \centering
    \begin{minipage}{0.7\textwidth}
        \centering
        \resizebox{\textwidth}{!}{
        \begin{tabular}{|c|c|c|c|c|c|}
        \hline
        Proximity Selection & Ngal & $\langle \sigma\rangle$ & $\sigma_{\langle \cos \theta \rangle}$ & $p_{KS}$\\
        \hline
        Spirals &  &   & &\\
         $d \le$ 0.2 Mpc & 3245 & 1.029 (Fig. \ref{Spiral - Function of distance}.1, red) & 0.025 (Fig. \ref{Spiral - Function of distance}.1A) & $7.3 \times 10^{-1}$\\
         0.2 $\le d \le$ 0.5 Mpc & 10144 & 0.499 (Fig. \ref{Spiral - Function of distance}.2, red) & 0.898 (Fig. \ref{Spiral - Function of distance}.2A) & $8.4 \times 10^{-1}$ \\
         $d \le$ 0.5 Mpc & 13389 & 0.660 (Fig. \ref{Spiral - Function of distance}.2, blue) & 0.742 (Fig. \ref{Spiral - Function of distance}.2B) & $9.1 \times 10^{-1}$ \\
         0.5 $\le d \le$ 1 Mpc & 9074 & 1.079 (Fig. \ref{Spiral - Function of distance}.3, red) & 1.919 (Fig. \ref{Spiral - Function of distance}.3A) & $3.9 \times 10^{-2}$ \\
         $d \le$ 1 Mpc & 22463 & 0.950 (Fig. \ref{Spiral - Function of distance}.3, blue) & 1.767 (Fig. \ref{Spiral - Function of distance}.3B) & $8.6 \times 10^{-2}$ \\
         1 $\le d \le$ 2 Mpc & 10054 & 0.945 (Fig. \ref{Spiral - Function of distance}.4, red) & 2.311 (Fig. \ref{Spiral - Function of distance}.4A) & $8.0 \times 10^{-3}$ \\
         $d \le$ 2 Mpc & 32517 & 1.176 (Fig. \ref{Spiral - Function of distance}.4, blue) & 2.795 (Fig. \ref{Spiral - Function of distance}.4B) & $1.6 \times 10^{-3}$ \\
        \hline
        Proximity Selection & Ngal & $\langle \sigma\rangle$ & $\sigma_{\langle \cos \theta \rangle}$ & $p_{KS}$\\
        \hline
        Ellipticals &  &  & &\\
         $d \le$ 0.2 Mpc & 1455 & 1.284 (Fig. \ref{Elliptical - Function of distance}.1, red) & 4.174 (Fig. \ref{Elliptical - Function of distance}.1A) & $5.7 \times 10^{-4}$ \\
         
         0.2 $\le d \le$ 0.5 Mpc & 4891 & 2.418 (Fig. \ref{Elliptical - Function of distance}.2, red) & 8.563 (Fig. \ref{Elliptical - Function of distance}.2A) & $1.5 \times 10^{-12}$ \\
         
         $d \le$ 0.5 Mpc & 6346 & 2.696 (Fig. \ref{Elliptical - Function of distance}.2, blue) & 9.500 (Fig. \ref{Elliptical - Function of distance}.2B) & $1.9 \times 10^{-14}$ \\
         
         0.5 $\le d \le$ 1 Mpc & 5751 & 1.766 (Fig. \ref{Elliptical - Function of distance}.3, red) & 6.989 (Fig. \ref{Elliptical - Function of distance}.3A) & $1.4 \times 10^{-7}$ \\
         
         $d \le$ 1 Mpc & 12097 & 3.107 (Fig. \ref{Elliptical - Function of distance}.3, blue) & 11.598 (Fig. \ref{Elliptical - Function of distance}.3B) & $3.6 \times 10^{-17}$ \\
         
         1 $\le d \le$ 2 Mpc & 6858 & 1.918 (Fig. \ref{Elliptical - Function of distance}.4, red) & 6.190 (Fig. \ref{Elliptical - Function of distance}.4A) & $3.9 \times 10^{-9}$ \\
         
         $d \le$ 2 Mpc & 18955 & 3.631 (Fig. \ref{Elliptical - Function of distance}.4, blue) & 13.067 (Fig. \ref{Elliptical - Function of distance}.4B) & $1.5 \times 10^{-27}$ \\
        \hline
        \end{tabular}
        }
        \tablefoot{Columns indicate: (1) Galaxy subsample defined by their distance from filament spine (in Mpc); (2) Number of galaxies in each subsample ($N_{gal}$); (3) Alignment signal Significance ($\langle \sigma \rangle$), quantifying deviation from randomized distributions; (4) Significance from Mean of Alignment signal($\sigma_{\langle \cos \theta \rangle}$), measuring deviation of observed mean from random expectations in standard deviations; and (5) Kolmogorov–Smirnov test probability ($p_{KS}$), evaluating the likelihood that observed alignment distributions observed from a random orientation scenario.}
        \label{TABLE1}
    \end{minipage}
\end{table*}

\section{Results}

In Fig. \ref{Alignment signal - Ellipticals and Spirals} we show the alignment for spiral (left) and elliptical (right) galaxies located within 2~Mpc of the filament spine. The upper row shows the probability distributions (solid line) and the result of the randomization procedure described above. The 1 $\&$ 2 $\sigma$ null hypothesis corridor is indicated by the dark and the light cyan bands. The reader will note that the randomization procedure does not produce a uniform band but rather imprints an alignment bias on the sample. The bottom rows of Fig. \ref{Alignment signal - Ellipticals and Spirals} show the exact same information but now normalized such that the randomization presents as uniform. Throughout the rest of the paper, we will display the alignment results in this way, the top row of Fig. \ref{Alignment signal - Ellipticals and Spirals} being presented for didactical reasons. Two immediate results of this plot stand out: 

1) Elliptical galaxies (right column) show a statistically significant perpendicular alignment between their short axes ($\hat{c}_{galaxy}$) and the filament spine ($\hat{L}_{filament}$). Alternatively put: their long axes align with the filament spine. It is detected at the $\sim$3.6 $\sigma$ level significance from the null hypothesis, and the mean of alignment signal shows a $\sim$13 $\sigma$ level significance from the distribution of mean from the null hypothesis cases. 

2) Only a statistically weak alignment ($\sim$1.2 $\sigma$) is seen when examining the full distribution of spiral galaxy angles. However, the mean angle of the distribution is $\sim$2.8 $\sigma$ inconsistent with the null hypothesis expectation. This is intriguing since 2.8$\sigma$ (99.47\% inconsistent with random) is approaching the threshold ($3\sigma$) most scientists would consider a statistically significant signal.

Taken together, Fig. \ref{Alignment signal - Ellipticals and Spirals} indicates that using the largest, most modern atlas of galaxy position angles and morphologies, together with the most expansive catalog of filaments, reaffirms the alignment between elliptical galaxies and the cosmic web and hints at a signal that is up to 99.5\% inconsistent with random for spirals.

As a natural progression, we extended our analysis to observe the alignment trend as a function of: 1) distance between the galaxies and the spine of their host filament $\&$ 2) absolute magnitude. Lastly, we also run a Machine Learning algorithm to isolate the properties that maximize the alignment signal, explained in Section \ref{Max.Subset}.

\subsection{Alignment signal as a function of filament proximity}

The alignment signal is examined as a function of distance from the filament spine across four discrete bins within 2 Mpc to see if any distance dominates the signal. The distances considered here are selected so as to compromise between having (roughly) the same number of galaxies per distance bin and physically relevant distance scales.
Fig. \ref{Spiral - Function of distance} and Fig. \ref{Elliptical - Function of distance} show how the alignment signal of the spiral and elliptical (respectively) depend on distance in both a cumulative (dashed) and differential (solid) fashion. 

Table \ref{TABLE1} presents the statistical analysis of galaxy-filament alignment for spiral and elliptical galaxies respectively. The table summarizes the following metrics: Mean of Alignment signal ($\langle \cos \theta \rangle$), Significance from Mean of Alignment signal ($\sigma_{\langle \cos \theta \rangle}$), Alignment signal Significance ($\langle \sigma\rangle$), and the Kolmogorov-Smirnov (KS) test p-value ($p_{KS}$) for different proximity selections.

\begin{figure*}[t]
    \centering
    \begin{minipage}{\textwidth}
        \centering
        \includegraphics[width=\textwidth]{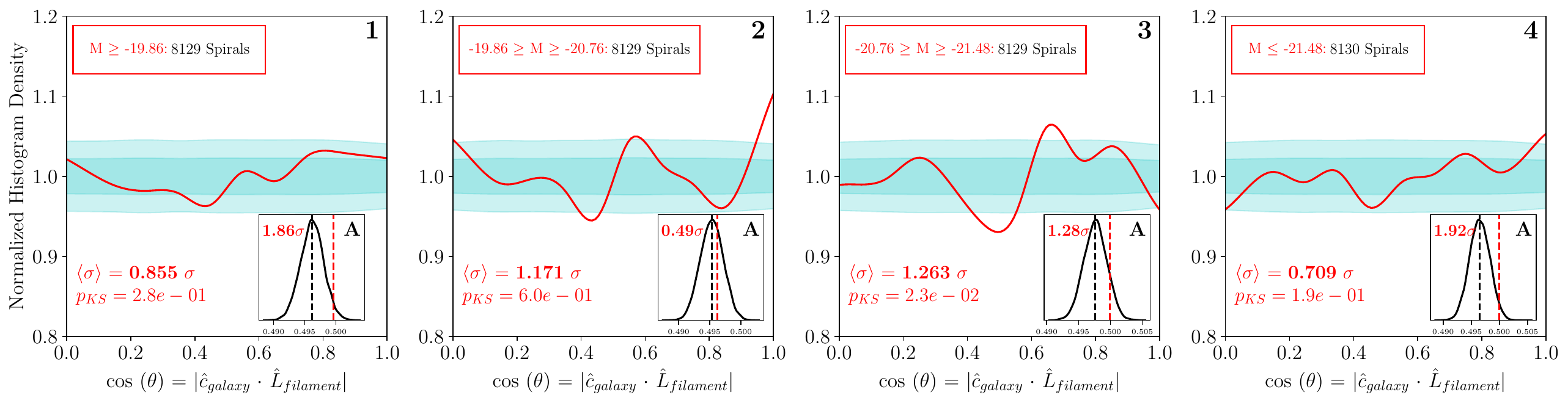}
        
        \caption{Normalized histogram of the alignment between spiral galaxy spins and cosmic filaments, based on different luminosity bins. Each panel represents the distribution of the alignment signal (cos ($\theta$)) for spiral galaxies with absolute magnitudes in the following ranges: (a) -19.86 $\le M$ (b) -20.76 $\le M \le$ -19.86 (c) -21.48 $\le M \le$ -20.76, and d) $M \le$ -21.48. The alignment is quantified as the cosine of the angle between the galaxy spin vector and the filament axis. An inset plot depicting the Mean of alignment signal (red line) along with the distribution of the mean from the null hypothesis cases is inserted to visualize how much is the mean of the alignment signal is further from the median of the distribution of the mean from the null hypothesis cases (black line), in terms of standard deviation of the distribution. The statistical significance of alignment in each subset is quantified and presented in Table \ref{TABLE2} (Note: The alignment signal are normalized with the mean of the random signal that the random is based around 1.0 for better inference of the alignment signal, and not under the assumption of null hypothesis to be an uniform distribution (refer Fig. \ref{Alignment signal - Ellipticals and Spirals}))}
        \label{Spirals - Function of luminosity}
    \end{minipage}
\end{figure*}

\begin{figure*}[t]
    \centering
    \begin{minipage}{\textwidth}
        \centering
        \includegraphics[width=\textwidth]{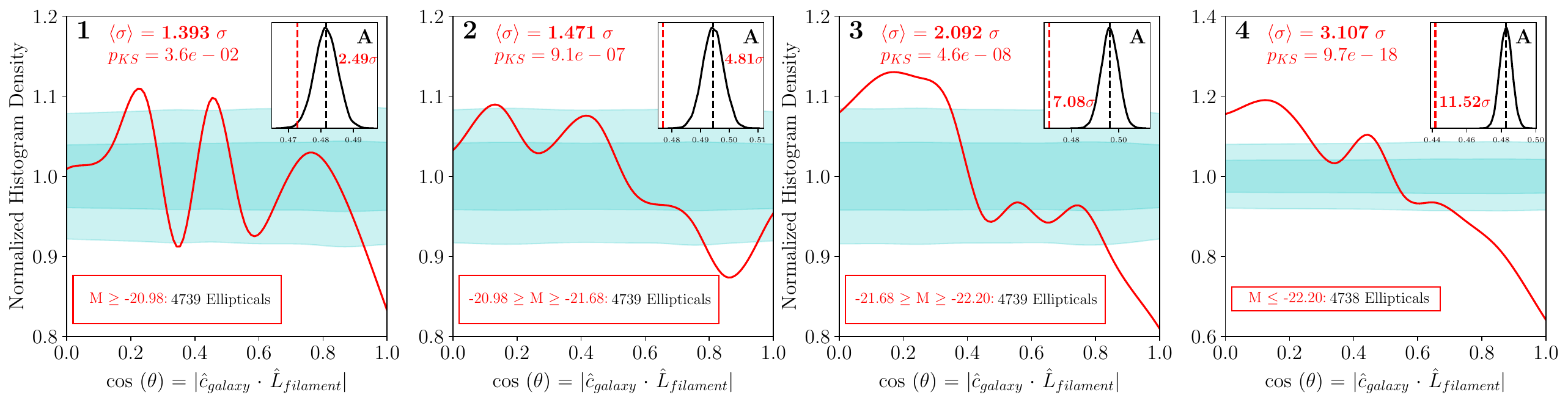}
        
        \caption{Normalized histogram of the alignment between elliptical galaxy spins and cosmic filaments, based on different luminosity bins. Each panel represents the distribution of the alignment signal (cos ($\theta$)) for elliptical galaxies with absolute magnitudes in the following ranges: (a) -20.98 $\le M$ (b) -21.68 $\le M \le$ -20.98 (c) -22.20 $\le M \le$ -21.68, and d) $M \le$ -22.20. The alignment is quantified as the cosine of the angle between the galaxy spin vector and the filament axis. An inset plot depicting the Mean of alignment signal (red line) along with the distribution of the mean from the null hypothesis cases is inserted to visualize how much is the mean of the alignment signal is further from the median of the distribution of the mean from the null hypothesis cases (black line), in terms of standard deviation of the distribution. The statistical significance of alignment in each subset is quantified and presented in Table \ref{TABLE2} (Note: The alignment signal are normalized with the mean of the random signal that the random is based around 1.0 for better inference of the alignment signal, and not under the assumption of null hypothesis to be an uniform distribution (refer Fig. \ref{Alignment signal - Ellipticals and Spirals}))}
        \label{Ellipticals - Function of luminosity}
    \end{minipage}
\end{figure*}

\begin{table*}[t]
    \centering
    \caption{Statistical analysis of galaxy-filament alignment signals for spiral and elliptical galaxies within various absolute luminosity (M) ranges}
    \begin{minipage}{0.7\textwidth}
        \centering
        \resizebox{\textwidth}{!}{
        \begin{tabular}{|c|c|c|c|c|}
        \hline
        Luminosity Selection & Ngal & $\langle \sigma\rangle$ & $\sigma_{\langle \cos \theta \rangle}$ & $p_{KS}$\\
        \hline
        Spirals &  &  & &\\
        -19.86 $\le M$ & 8129 & 0.855 (Fig. \ref{Spirals - Function of luminosity}.1)  & 1.856 (Fig. \ref{Spirals - Function of luminosity}.1A) & $2.8 \times 10^{-1}$\\
        -20.76 $\le M <$ -19.86 & 8129 & 1.171 (Fig. \ref{Spirals - Function of luminosity}.2) & 0.491 (Fig. \ref{Spirals - Function of luminosity}.2A) & $6.0 \times 10^{-1}$\\
        -21.48 $\le M <$ -20.76 & 8129 & 1.263 (Fig. \ref{Spirals - Function of luminosity}.3) & 1.283 (Fig. \ref{Spirals - Function of luminosity}.3A) & $2.3 \times 10^{-2}$\\
        $M <$ -21.48 & 8130 & 0.709 (Fig. \ref{Spirals - Function of luminosity}.4) & 1.924 (Fig. \ref{Spirals - Function of luminosity}.4A) & $1.9 \times 10^{-1}$\\
        \hline
        Ellipticals &  &  & &\\
        -20.99 $\le M$ & 4739 & 1.393 (Fig. \ref{Ellipticals - Function of luminosity}.1)  & 2.492 (Fig. \ref{Ellipticals - Function of luminosity}.1A) & $3.6 \times 10^{-2}$ \\
        -21.69 $\le M <$ -20.99 & 4739 & 1.471 (Fig. \ref{Ellipticals - Function of luminosity}.2) & 4.809 (Fig. \ref{Ellipticals - Function of luminosity}.2A) & $9.1 \times 10^{-7}$ \\
        -22.20 $\le M <$ -21.69 & 4739 & 2.092 (Fig. \ref{Ellipticals - Function of luminosity}.3) & 7.084 (Fig. \ref{Ellipticals - Function of luminosity}.3A) & $4.6 \times 10^{-8}$ \\
        $M <$ -22.20 & 4738 & 3.107 (Fig. \ref{Ellipticals - Function of luminosity}.4) & 11.524 (Fig. \ref{Ellipticals - Function of luminosity}.4A) & $9.7 \times 10^{-18}$ \\
        \hline
        \end{tabular}
        }
        \tablefoot{Columns indicate: (1) Luminosity selection range (in absolute magnitude); (2) Number of galaxies in each bin ($N_{gal}$); (3) Alignment signal Significance ($\langle \sigma \rangle$), indicating the overall alignment strength; (4) Significance from Mean of Alignment signal ($\sigma_{\langle \cos \theta \rangle}$), quantifying the deviation of the observed alignment from random expectations in standard deviation units; and (5) Kolmogorov–Smirnov test p-value ($p_{KS}$), assessing the likelihood of the alignment signal arising by chance.}
        \label{TABLE2}
    \end{minipage}
\end{table*}

We first turn to Fig. \ref{Spiral - Function of distance}, the alignment of spiral galaxies with filaments. Here, the results are unambiguous. At no filamentary distance is the full distribution of the galaxies significantly aligned with the cosmic web in terms of the Alignment signal Significance ($\langle \sigma\rangle$). The cumulative distributions (dashed blue lines) are well within their (lighter blue dot-dashed lines) error corridors, as are the solid lines. From Table \ref{TABLE1}, the significance of alignments are all either well under or slightly above 1 sigma. 

The mean of the alignment signal (cos $\theta$) for both the cumulative and the differential subsets are also centered around 0.5, hinting that the disk galaxies used in this analysis don't show a significant alignment trend, irrespective of their proximity (also supported by negligible values of $p_{KS}$). We note that there is a small exception - in the $0.5<d<1.0$ Mpc and $1.0<d<2.0$ Mpc range, the roughly 9,000 and 10,000 spirals in each distance bracket have a mean angle that is $1.9\sigma$ and $2.3\sigma$ inconsistent with random. Although the full distributions remain consistent with random at the $\sim 1\sigma$ level, it is noteworthy that the mean behaves somewhat atypically. 

In Fig. \ref{Elliptical - Function of distance} the alignment for ellipticals is shown, also as a function of filamentary distance. Here, the plots indicate that in the inner core of the filament the alignment is inconsistent with random at a mere 1.2 $\sigma$ level. Beyond 0.2 Mpc, more significant alignments emerge. The differential lines (solid) indicate a mixed picture with statistically significant alignments peaking at the 0.2 - 0.5 Mpc cylindrical annulus ($2.42\sigma$), slightly decreasing in the 0.5 - 1 Mpc range ($1.8\sigma$) and then rising again in the outskirts ($1.9\sigma$). 

The cumulative (solid) lines maintain a consistent trend of increasing significance, as more galaxies are included. Table \ref{TABLE1} also shows that other significance indicators show consistent trends with alignment signal significance ($\langle \sigma \rangle$) that the alignment is statistically significant for the subset of elliptical galaxies between 0.2 - 0.5 Mpc from the spine of the filament. Contrary to spirals, the $\langle \cos \theta \rangle$ values for each subset of elliptical galaxies from Table \ref{TABLE1} are consistently smaller than the median of the distribution of the mean from the null hypothesis case (see inset of Fig. \ref{Elliptical - Function of distance}) indicating further that the ellipticals indeed have a preferred orientation (namely perpendicular to the filament spine) as a function of proximity to the filament, which is confirmed by larger $\sigma_{\langle \cos \theta \rangle}$ significance and significant $p_{KS}$ values (Table \ref{TABLE1}).

\subsection{Alignment signal as a function of absolute luminosity}

We attempted to study the alignment signal as a function of brightness, since theoretical studies such as \citet{Lee_2004} found that brighter spiral galaxies had a more pronounced alignment with the filaments compared to their lesser luminous counterparts. Here, in this study, we have performed our analysis as a function of absolute luminosity in the R band for both spiral and elliptical galaxies.

Fig. \ref{Spirals - Function of luminosity} and Fig. \ref{Ellipticals - Function of luminosity} show the alignment trend for both spiral and elliptical galaxies across four luminosity bins, each containing $\sim$ 8130 spirals and $\sim$ 4739 ellipticals, respectively. For spirals (Fig. \ref{Spirals - Function of luminosity}, Table \ref{TABLE2}), none of the subsets exhibit a strong alignment. The most pronounced signal, in terms of significance ($\langle \sigma \rangle$) occurs in the third bin (\(-21.48 \le M < -20.76\)), where the mean alignment strength rises to \(1.263\,\sigma\) above the randomized expectation, with a $p_{\rm KS}$ value of $2.3\times10^{-2}$.

However, even here the absolute cosine \(\langle\cos\theta\rangle \simeq 0.500\) remains essentially indistinguishable from random, matching well with the visual inspection of the alignment signal, which appears to be random. On the other hand, the brightest disk galaxies show an alignment trend and are additionally supported by a comparatively higher $\sigma_{\langle \cos \theta \rangle}$, again has a \(\langle\cos\theta\rangle \simeq 0.500\), indicating a very weak alignment trend (also inferred from their respective $p_{KS}$ \& $\langle \sigma \rangle$ values).

By contrast, ellipticals (Fig. \ref{Ellipticals - Function of luminosity}, and again Table \ref{TABLE2}) display a steadily strengthening perpendicular alignment as one moves to brighter bins.  The faintest ellipticals (\(M \ge -20.99\)) already show a marginal \(1.393\,\sigma\) significance (\(p_{\rm KS}=3.6\times10^{-2}\)), but this rises to \(1.471\,\sigma\) (\(p_{\rm KS}=9.1\times10^{-7}\)) for \(-21.69 \le M < -20.99\), to \(2.092\,\sigma\) (\(p_{\rm KS}=4.6\times10^{-8}\)) for \(-22.20 \le M < -21.69\), and peaks at \(3.107\,\sigma\) (\(p_{\rm KS}=9.7\times10^{-18}\)) for the brightest ellipticals (\(M < -22.20\)). 

Thus, massive ellipticals exhibit the strongest and most significant perpendicular alignment with the filamentary network, whereas spiral disks show no analogous luminosity trend.

\subsection{Obtaining the subset with maximum significance}\label{Max.Subset}
Observational studies have consistently shown that recovering alignment signal with spirals is only possible under stringent sub-sample cuts, e.g. selecting bright spirals \citep{Tempel2013}, pure disk or low bulge-fraction systems \citep{Barsanti:2022}, high-spin late types (fast rotators) \citep{Kraljic:2021}, or edge-on disks \citep{Jones2010}. To further refine our sensitivity to the notoriously weak spiral–filament alignment and to see if there is a specific subset with maximum statistical significance, we implemented a Bayesian optimization involving the two factors we performed our study on: Absolute \(r\)–band magnitude and projected filament distance.

\begin{table*}[t]
    \centering
    \caption{Statistical analysis of subsample from spiral and elliptical galaxies as function of proximity and luminosity with significant alignment}
    \begin{minipage}{0.8\textwidth}
        \centering
        \resizebox{\textwidth}{!}{
        \begin{tabular}{|c|c|c|c|c|c|}
        \hline
        Proximity Selection & Luminosity Selection & Ngal & $\langle \sigma\rangle$ & $\sigma_{\langle \cos \theta \rangle}$ & $p_{KS}$\\
        \hline
        Spirals &  &  &  & &\\
        0.36 $\le d \le$ 2 Mpc & -23.46 $\le M \le$ -16 & 23163 & 1.567 (Fig. \ref{BestFit}.1) & 3.859 (Fig. \ref{BestFit}.1A) & $4.2 \times 10^{-5}$\\
        \hline
        Ellipticals &  & &  & &\\
        0 $\le d \le$ 2 Mpc & -24 $\le M \le$ -20.67 & 15343 & 3.772 (Fig. \ref{BestFit}.1) & 13.792 (Fig. \ref{BestFit}.2A) & $6.6 \times 10^{-28}$ \\
        \hline
        \end{tabular}
        }
        \tablefoot{Columns indicate: (1) Proximity selection range (in Mpc) from the filament spine; (2) Luminosity selection range (in absolute magnitude); (3) Number of galaxies ($N_{gal}$); (4) Alignment signal Significance ($\langle \sigma \rangle$), indicating the overall alignment strength; (5) Significance from Mean of Alignment signal ($\sigma_{\langle \cos \theta \rangle}$), quantifying the deviation of the observed alignment from random expectations in standard deviation units; and (6) Kolmogorov–Smirnov test p-value ($p_{KS}$), assessing the likelihood of the alignment signal arising by chance.}
        \label{TABLE3}
    \end{minipage}
\end{table*}

We employed a Gaussian process (GP)–based Bayesian optimization method (using the \texttt{gp\_minimize} routine from \textsc{scikit-optimize}). This approach constructs a probabilistic surrogate model of the objective function to efficiently explore a four-dimensional parameter space—defined by the minimum and maximum absolute $r$-band magnitudes ($M_{r}^{\min},M_{r}^{\max}$) and the minimum and maximum projected filament distances ($d_{\rm fil}^{\min},d_{\rm fil}^{\max}$)—without requiring an exhaustive grid search. 

We performed the optimization separately for the spiral and elliptical samples, allowing each morphological class to have distinct optimal criteria. For spirals, the algorithm converged on subsets of relatively bright disk galaxies at intermediate-to-large filament distances (approximately $-23.5 \le M_{r} \le -16.0$ and 0.36 $\le d_{\rm}\le$ 2.00 Mpc), whereas for ellipticals the highest significance was achieved by focusing on the brightest ellipticals ($M_{r}\lesssim -20.7$) over the full distance range ($d\le2.0$\,Mpc). The Bayesian optimizer inherently penalizes overly restrictive cuts that yield insufficient sample sizes by returning low objective values when the subset size falls below the reliability threshold, thus ensuring that the resulting subsets both maximize the alignment signal and retain adequate statistical robustness. Even after this targeted selection, spiral alignments remain marginal.

\begin{figure}
    \begin{center}
    \includegraphics[width=\textwidth/2]{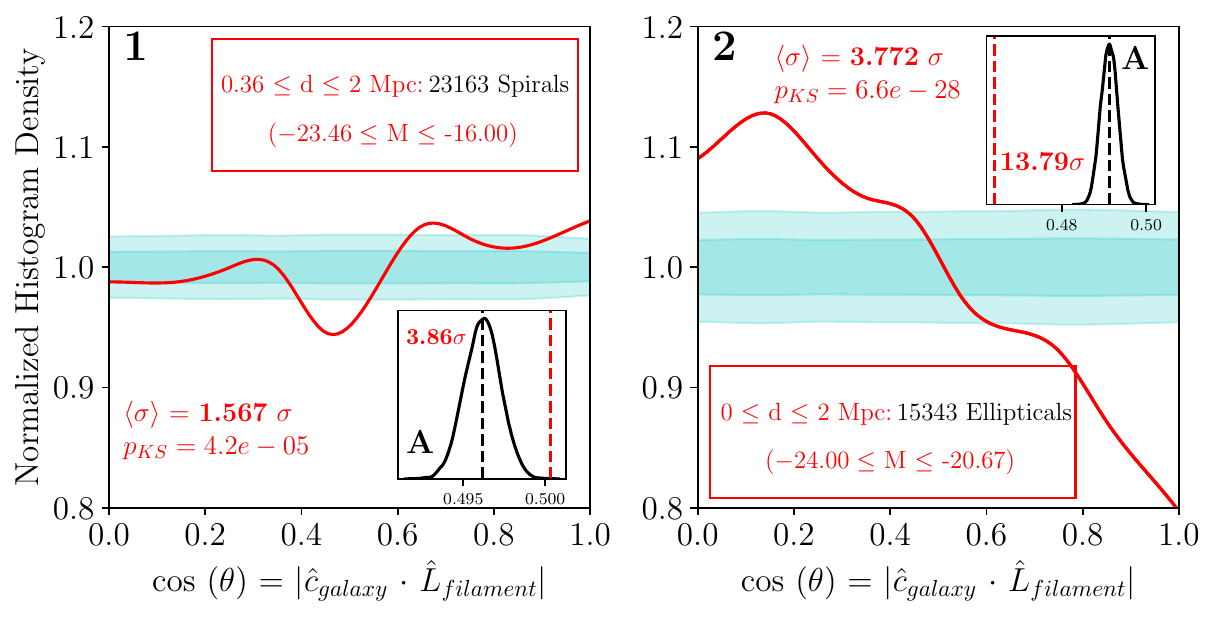}
    \caption{Normalized histogram of cosines of the angles between spines of the filaments and the spin axes of subset of spiral and elliptical galaxies with maximum significance. (\( \cos \theta = |\hat{c}_{\text{galaxy}} \cdot \hat{L}_{\text{filament}}| \)). The dark cyan band represents the 1 $\sigma$ null hypothesis corridor, beyond which the probability density function is deemed significant, while the light cyan band provides a 2$\sigma$ corridor as a visual comparison. An inset plot depicting the Mean of alignment signal (red line) along with the distribution of the mean from the null hypothesis cases is inserted to visualize how much is the mean of the alignment signal is further from the median of the distribution of the mean from the null hypothesis cases (black line), in terms of standard deviation of the distribution. The statistical significance of alignment of these
subset with maximum significance is quantified and presented in Table \ref{TABLE3}}
    \label{BestFit}
    \end{center}
\end{figure}

Fig. \ref{BestFit} provides the alignment trend for the subset of early type and disk galaxies with maximum statistical significance. Table \ref{TABLE3} summarizes the respective statistical metrics for these subsets. Even after this targeted selection, the spiral subset attains only a marginal \(\sim1.6\sigma\) signal, consistent with previous findings that broadly defined spiral samples mask the expected parallel alignment unless one isolates the most disk-dominated, well-inclined galaxies \citep[for e.g.][]{Hirv2017, Pahwa2016}. ($\sim1.6\sigma$), while ellipticals exhibit a robust $\sim3.8\sigma$ perpendicular alignment. Note that when the mean value is examined, the spirals are 3.8$\sigma$ away from random and the ellipticals 13.8$\sigma$ events. This application of Bayesian optimization therefore pinpoints the luminosity and environmental regimes within the galaxy catalog in which the cosmic spin–filament alignment is most pronounced.

\section{Summary and conclusion}

In this paper we have examined the alignment between spiral and elliptical galaxies and the cosmic network of filaments. A large body of work has been published dedicated to examining such an alignment: whether it is to understand the origin of angular momentum \citep{schafer2009} or intrinsic alignments as a nuisance parameter in lensing surveys \citep{Amon:2022}, the positioning of galaxies is a well-studied phenomenon in observations as well as simulations. Here we repeat previous studies with the largest sample of elliptical and spiral galaxies to date.

A number of challenges -- or partially arbitrary choices -- exist that influence the detection of a potential signal. The first set of questions concerns the method of quantifying the Large Scale Structures (LSS) (e.g. \citet{Libeskind2018}). Here we opt for a fairly straightforward method, the point process known as the Bisous method, whose direction has been shown to be consistent with, if not all, at least 2 methods -- the V and T web, \citet{Hoffman2012} and \citet{Hahn2010} respectively \citep[see][]{2015MNRAS.453L.108L}. The identification of the LSS is not as ambiguous or degenerate so as to cause major concern; however, it is a source of some uncertainty.

More critically is the determination of the spin axis of a galaxy. As written above, we must transform a projected ellipse into a 3D ellipsoid. Multiple degeneracies exist that all serve to wipe out an intrinsic signal among them, the handedness and the line of sight degeneracy, namely if an axis is pointing towards or away from the observer at some inclination angle. While employing galaxy modeling as performed here (e.g. \citet{Lee2007}) helps to strengthen any potential signal, it's clearly a small step in the right direction. Determining the true 3D short axis of a galaxy remains the greatest unknown in works such as this.

Given all the issues mentioned above, it is remarkable that any signal persists.  Ellipticals exhibit a strong perpendicular alignment: in the optimal bin ($0\le d\le2\,$Mpc, $-24\le M\le -20.7$) we measure 
$\langle\sigma\rangle =$ 3.772 $\sigma$, $\sigma_{\langle \cos \theta \rangle} =$  13.792 $\sigma$ and the $p_{\rm KS}=6.6\times10^{-28}$, 
consistent with both earlier observations \citep[e.g.][]{TempelLibeskind2013} and hydrodynamical simulations of massive, pressure‐supported galaxies \citep[e.g.][]{Codis2018,Kraljic2020}.  The slight variation of signal strength with filament proximity may reflect filament boundary effects \citep{WangLibeskind2020}.

Spirals, by contrast, show no alignment in the full SGA sample, but a targeted selection ($0.36\le d\le2\,$Mpc, $-23.46\le M\le-16$) reveals a faint {\it parallel} signal when the full distribution is considered (  
$\langle\sigma\rangle =$ 1.567 $\sigma$) but a stronger signal when examining the median of the distribution ($\sigma_{\langle \cos \theta \rangle} =$  3.859 $\sigma$ and the $p_{\rm KS}=6.6\times10^{-28}$, Fig. \ref{BestFit}, Table~\ref{TABLE3}).  This echoes recent results from MaNGA \citep{Kraljic:2021}, the SIMBA simulation \citep{Kraljic2020}, and the CHILES/Blue Bird HI survey \citep{BlueBird:2020}, all of which report low‐mass systems’ spins parallel to filaments, which is a trend also observed in early studies like \citep[for e.g.][]{TempelLibeskind2013,Codis2018}

A note is in order regarding the difference between the statistical significances found when comparing the full distribution and the median values to random samples. It may be counterintuitive that these metrics disagree. Yet, the reader may consider the following. The median value is not sensitive to individual outliers, but is sensitive to groups of outliers. While the full distribution of the measured signal may well lie within the variance of the random distributions, its median is far more sensitive to outlying values. Ultimately, we only have one measured median, and we are comparing this to 10,000 random medians. The standard deviation of these 10,000 random medians probes the likelihood that one would obtain as many outliers as the signal. This is shown to be a far more sensitive metric than comparing the full distributions - and we note that a sensitive metric is precisely what is needed when trying to suss out a weak signal, buried in statistical noise, and compromised by inclination and shape degeneracies.

The faintness of the SGA spiral signal likely stems from the aforementioned effects (inclination, degeneracies, etc.), and on the other hand, it could also be possible misalignment between stellar disks and their host dark halos \citep[e.g.][]{SAMI_BlandHawthorn_2020}. More studies (specifically assessing the spin direction of satellites within a host) could shed light on the faint alignment detection with spirals and also enable us to pin down the selection criteria that would give us the subset of disk galaxies with a profound alignment signal and understand the principle behind it.

In the coming years, new observational data will allow us to study the intrinsic connection between galaxies and the cosmic web with great detail. One of the most promising surveys are the 4MOST \citep{2019Msngr.175....3D} WAVES \citep{Driver:2019} and 4HS \citep{2023Msngr.190...46T} surveys that will map the cosmic web in the nearby universe ($z$ < 0.2) with great detail down to the megaparsec scale, allowing us to see the small filaments and tendrils that influence the alignment of galaxies in the cosmic web.

\begin{acknowledgements}
We acknowledge the funding from the European Union's Horizon Europe research and innovation programme (EXCOSM, grant No. 101159513). ET was funded by the Estonian Ministry of Education and Research (grant TK202), Estonian Research Council (grant PRG1006). 
\end{acknowledgements}

%\bibliographystyle{aa}
%\bibliography{aa54113_25}

\end{document}